\documentclass{postunc}
\usepackage{tikz}
\usepackage{amsmath}
\usepackage{cleveref}
\usepackage{parskip}
\usepackage[square,numbers,sort&compress]{natbib}
\usepackage[charter]{mathdesign}
\usepackage{microtype}
\usepackage{xcolor}

\newcommand{\doi}[1]{\href{https://doi.org/#1}{https://doi.org/#1}}

\title{A HIERARCHICAL BAYESIAN APPROACH FOR POPULATION-BASED SHM IN SHIP HULL
STRUCTURES}

\author{Georgios Aravanis$^1$, Nicholas Silionis$^2$, Jacopo Bardiani$^1$, Marco Giglio$^1$, Konstantinos Anyfantis$^2$, and Claudio Sbarufatti$^1$}

\heading{G.\ Aravanis et al.}

\address{$^1$Politecnico di Milano, Department of Mechanical Engineering \\
  Via La Masa 1, 20156 Milano, Italy\\
  e-mail: \{georgios.aravanis, jacopo.bardiani, marco.giglio, claudio.sbarufatti\}@polimi.it \and
  $^2$
  National Technical University of Athens, School of Naval Architecture and Marine Engineering \\
  9 Heroon Polytechniou Av., Athens, 15780 Zografos, Greece\\
  e-mail: \{nsil,kanyf\}@naval.ntua.gr}


\abstract{Structural health monitoring (SHM) strategies involve the processing of structural response data to indirectly assess an asset’s condition. These strategies can be enhanced for a group of structures, especially when they are similar, since mutual underlying physics are expected to exist. The concept behind population-based SHM exploits the sharing of data among individuals, so that data-rich members can support data-scarce ones. One approach to population-level modeling is the hierarchical Bayesian method, where the model is structured hierarchically in terms of its parameters, and correlation among learning
tasks is enabled by conditioning on shared latent variables. 

This work investigates the application of a hierarchical Bayesian model to infer expected distributions of deflection amplitudes at both the population
and domain levels, with the aim of detecting excessive initial deflections in a population of plate elements. Although these damages are typically localized, they can trigger unexpected events, if not properly monitored.
The work is conducted in a numerical setting using a Finite Element model to generate strain response data, which serve as the monitoring data. Bayesian inference was conducted using Markov Chain Monte Carlo (MCMC), with a surrogate model employed to calculate the likelihood function. The hierarchical approach was compared to an independent model for a plate component with few data. The results revealed that, under data sparsity conditions, the hierarchical model can offer more robust results in terms of uncertainty, which is essential for decision-making tasks.}

\begin{document}

\maketitle

\section{INTRODUCTION}

Oftentimes, failure of many engineering assets is associated with localized damage events. %
In the case of ship hull structures, which are primarily composed of plates and stiffened panels, deterioration of these components can lead to reduced load-carrying capacity and potentially unexpected failure modes. %
Thus, the deployment of Structural Health Monitoring (SHM) campaigns on-board~\cite{farrar2012, farrar2006}, which are able to monitor such events, becomes highly relevant for helping operators make timely maintenance decisions.

In SHM, the challenge essentially lies in learning Quantities of Interest (QoIs) that characterize the underlying deterioration process. %
These are most often unobservable, and thus structural response quantities (e.g., strain or acceleration sensor data) are instead used to determine their values. %
Concurrently, this inference procedure needs to be robust in terms of uncertainty, as the confidence level in the deployed SHM system for decision-making directly depends on how well uncertainty is handled. %
This problem formulation lends itself well to a Bayesian framework~ \cite{gelman_2013}, where the goal is to estimate the updated joint probability distribution of the QoIs conditioned on the acquired sensor data. %
Acknowledging this, numerous studies in the SHM field have employed Bayesian methods to address different tasks in the classical SHM hierarchy~\cite{rytter_1993}; namely, related to (damage) diagnosis~\cite{gardner_2022}, identification~\cite{ramancha_2022}, and prognosis~\cite{cristiani_panel_2021, cristiani_coupons_2021}.

Despite the benefits of developing SHM approaches for structural integrity management purposes, their effectiveness is bounded by the availability of data relevant to the problem at hand. %
In real-world large-scale structures, this poses a critical challenge~\cite{cawley_2018}, since increased resource restrictions arising from the scale of the system influence data availability. %
At the same time, their inherent complexity requires sufficiently comprehensive datasets to describe the various phenomena, which one seeks to model. %
On the other hand, such structures often exhibit fundamental similarities in terms of their constituent components, presenting an opportunity to enable information sharing among them. %
If we know that they have been produced under similar specifications and with the same material, they can be considered nominally identical~\cite{bull_2021}. %
Even then however, they are not truly identical due to inherently random events (i.e., relating to manufacturing, assembly, materials, etc.). %
An example of this is the repeated presence of plate elements in a ship hull. %
Although produced on the same manufacturing line, they exhibit geometric variability due to fabrication-related factors during the ship’s construction, which often manifests as deviations in their intended shape.

The variation in the structural attributes of individual components results in variations in their response, which complicates the development of a model that generalizes well across all of them. %
Hierarchical Bayesian modeling provides a principled statistical framework to model these variations within a group while leveraging statistical similarities among its members~\cite{gelman_2007, gkreft_1998}, all while providing the key benefits of Bayesian modeling. %
The statistical correlations enabled through hierarchical models allow for improved inference in sparse data scenarios, whether due to data loss (e.g. transmission issues, sensor failure, etc.) or due to the challenges of online monitoring, where early predictions must be made with limited initial observations. %
An additional benefit of this approach is that it makes the most out of well-instrumented domains, and thus the overall added value of SHM implementation is increased compared to a more standard approach.

Hierarchical Bayesian models have seen particular success in several works related to population-based SHM~(PBSHM)~\cite{bull_2023, dardeno_2024, brealy_2024}, as well as other engineering problems framed in a hierarchical manner \cite{papadimas_2021, francesco_2021}. %
This is because of their ability to 
capture both common trends across a population as well as individual differences among its members, which aligns naturally with the concept of constructing a general population model. %
However, despite their benefits, these models have yet to be explored in SHM applications for marine structures.

In this work, we demonstrate a hierarchical Bayesian framework on a
large-scale double-bottom model of a containership, where damage is defined as a set of bathtub-shaped plate deflections. %
The approach involves first generating realistic strain observations by Finite Element~(FE) modeling. %
These observations are subsequently incorporated into a hierarchical model, alongside an FE-based surrogate, necessary for the calculation of the likelihood function and, ultimately, the prediction of parameters associated with the deflection amplitudes. %
Results are presented in terms of variance in the parameter estimates to assess whether data-poor plates can benefit from statistical information at the population level, to enable robust damage detection under data-sparsity.

\section{THE DATASET}\label{s:Data}

The dataset used in this work was generated from a detailed FE model, which corresponds to part of the double-bottom structure within the central hold of a containership, and is of size 27 m $\times$ 18 m $\times$ 2.2 m. %
A detailed description of the structural information has been omitted here for the sake of brevity but can be found in~\cite{aravanis_2023}. %
This section of the ship hull was selected because it is typically under constant in-plane compressive loading due to the nature of the operational loads a containership experiences. %
The combination of compression and initial out-of-plane deflections for plate elements can accelerate the onset of buckling, which makes this region particularly critical.

\subsection{Finite element modeling particulars}

\begin{figure}[ht]
  \begin{center}
    \includegraphics[width=11.5cm]{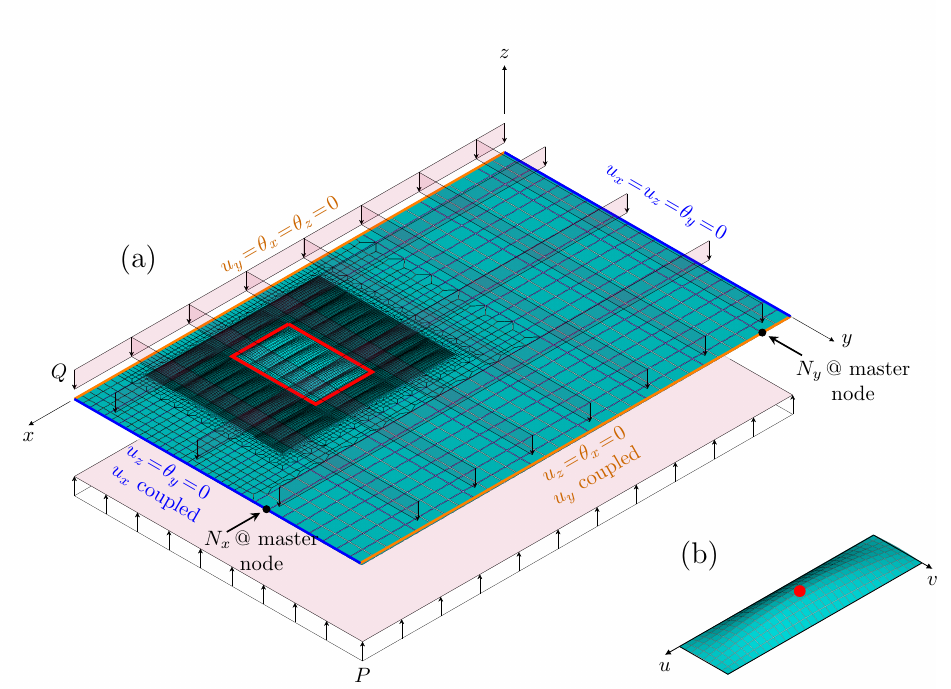}
    \caption{(a) FE model of the subject geometry including the monitored plates (within red boundaries), and (b) representative out-of-plane deflection with strain sensor at the center. Adapted from the original work of the authors~\cite{aravanis_2023}.}
    \label{fig:FEM}
  \end{center}
\end{figure}

The FE model features 4-node linear shell elements for the external bottom and 2-node linear beam elements to represent the various stiffening members of the double-bottom (i.e., girders, floors, longitudinal stiffeners). The stiffness contributed by the inner bottom has been implicitly captured by incorporating an effective width for the girders and floors. The group of $K=6$ plates enclosed within the red boundaries in \Cref{fig:FEM}(a) defines the region used for strain data generation for the purposes of this study. Each plate, of size $a \times b$ , deforms based on a sinusoidal expression given by:

\begin{equation}
  w(u,v;\tilde{w}) = \tilde{w}  \sin \left(\frac{\pi u}{a}\right) \sin \left(\frac{\pi v}{b}\right)
\end{equation}

\noindent where $\tilde{w}$ denotes the amplitude of the deflection, $u \in [0,a]$ and $v \in [0,b]$. %
Their adjacent plates were also modeled as deflected to provide realistic states for the surroundings of the monitored region, since all structural elements are expected to exhibit slight imperfections within allowable limits. %
They are shown with a slight black shading to indicate that they are not part of the data generation process. %
During the modeling process, $\tilde{w}$ was controlled parametrically to generate different realizations of out-of-plane deflections. %
The perspective shape of a simulated out-of-plane deflection can be seen in \Cref{fig:FEM}(b), where the (conceptual) strain sensor location is also indicated by a red marker. %
This placement is consistent across all $K$ plates.

The loads considered are those typically experienced by a containership while sailing in calm water conditions; namely, hogging (i.e., negative) hull girder bending, lateral pressure on the side shell and outer bottom, and the cargo weight \cite{sumi_2015}. %
At the double-bottom level, these translate into 
longitudinal ($N_{{x}}$) and transverse ($N_{{y}}$) in-plane compressive stresses, pressure $(P)$ distributed over the hull exterior, and line loads $(Q)$ on the inner bottom, respectively. %
These were assigned with representative values based on the loading manual of the vessel. %
In terms of the boundary conditions, they were defined based on how the considered geometry deforms with respect to the entire structure and are detailed in the descriptions adjacent to the edges of the model.

Regarding the material, linear elastic and isotropic properties (Young’s modulus $E = 207$~GPa, Poisson’s ratio $v = 0.3$) were assigned. %
A static, linear elastic analysis was ultimately employed to solve the model, essentially representing conditions when the vessel is in port or sailing in still water (i.e., no waves). %
Nevertheless, the data sampling rate implied in this context is considered sufficient, given the relatively low temporal evolution of out-of-plane deflections.

\subsection{Dataset generation}

In the data generation process, transverse strains $\varepsilon_{yy}$ were selected as the observed quantities, which falls in line with the underlying physics of the problem, as plate bending theory dictates that strains in the shorter dimension are the dominant ones. %
Hereinafter, these are denoted as $\varepsilon$ for notational simplicity.

To generate the latent quantities $\tilde{w}$ that give rise to these observations, it is assumed that each plate $k \in K$ is deflected based on a local (Normal) distribution for $\tilde{w}$, whose parameters are themselves realizations from common higher-level distributions. %
To this end, global distributions were placed over both the expected value of the deflection amplitude and its variance. %
The former captures the variability introduced by fabrication processes, leading to differences in expected deflection amplitudes among plates. %
Meanwhile, the latter accounts for changing conditions that introduce additional variability around each expected deflection, reflecting uncertainties in the structural response beyond systematic fabrication-induced deviations. 

\begin{figure}[ht]
  \begin{center}
    \includegraphics[width=10.0cm]{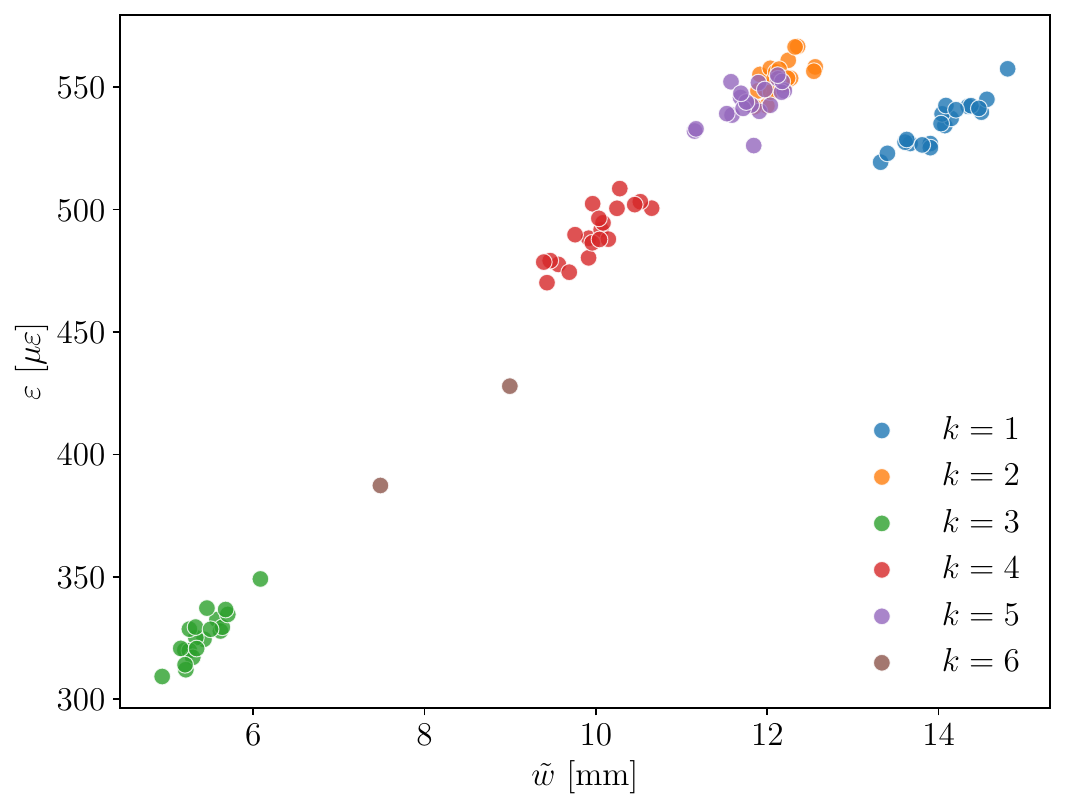}
    \caption{Clusters of generated observations of transverse strain for the six plate components. The brown markers show Plate 6 with scarce data.}
    \label{fig:clusters}
  \end{center}
\end{figure}

By sampling from these global distributions, $K$ pairs of local expectations and variances were obtained, defining an equal number of local distributions, from each of which 
$N_k$ samples of $\tilde{w}$ were drawn. %
The resulting matrix of $K$-dimensional vectors $\left\{ \left[\tilde{w}_{i,1}, \ldots, \tilde{w}_{i,k}, \dots, \tilde{w}_{i,K} \right] \right\}_{i=1}^{N_k} \in \mathbb{R}^{N_{k} \times K}$ was concatenated with additional samples corresponding to the deflections of the neighboring plates, which were assumed to remain within allowable tolerances. %
The augmented vectors were fed into the constructed FE model to generate transverse strain observations $\varepsilon$, which were subsequently perturbed with additive white Gaussian noise with a standard deviation of $5 \; \mu \varepsilon$. %
The corresponding
pairs of $\left\{\tilde{w}_{i,k}, \varepsilon_{i,k} \right\}_{i=1}^{N_k}$ for each $k \in K$ are shown with different colors in \Cref{fig:clusters}, where $N_{k} = 20$ for $k \in \{1,\ldots,5\}$ and $N_{k} = 2$ for $k=6$. %
In this way, some measurements are hidden from the employed models, to demonstrate the robustness of hierarchical modeling in imbalanced dataset settings.

\section{THE HIERARCHICAL BAYESIAN APPROACH}\label{s:HBM}

To establish the hierarchical Bayesian formulation, we represent the data from the entire group of plates as:
\begin{equation}
    \left\{\mathbf{\tilde{w}}_{k}, \boldsymbol{\varepsilon}_{k} \right\}_{k=1}^K = \left\{\left\{\tilde{w}_{i,k}, \varepsilon_{i,k} \right\}_{i=1}^{N_k}\right\}_{k=1}^K
\end{equation}

The objective is to learn the posterior distributions of the parameters (local expectations and variances) and hyperparameters (higher-level expectations and variances) that govern the deflection amplitude $\tilde{w}$, given the observed strain data $\varepsilon$.

\subsection{Surrogate modeling}

Each plate (domain) is related to its own task, which can be formally expressed as:
\begin{equation}
    \varepsilon_{i,k} = \mathcal{M}(\tilde{w}_{i,k}) + \epsilon_{i,k}
    \label{eq:3}
\end{equation}

\noindent where $\mathcal{M}(\cdot)$ denotes the model operator and $\epsilon_{i,k}$ is the additive noise term. %
The former represents the expected mapping from deflections to transverse strain, established by the constructed FE model, which was eventually replaced with a cheaper data-driven surrogate. 
This is typical practice to avoid computational overhead when employing sampling-based algorithms (i.e., Markov Chain Monte Carlo (MCMC)) for the estimation of posterior distributions in a Bayesian context.

\begin{figure}[ht]
  \begin{center}
    \includegraphics[width=10cm]{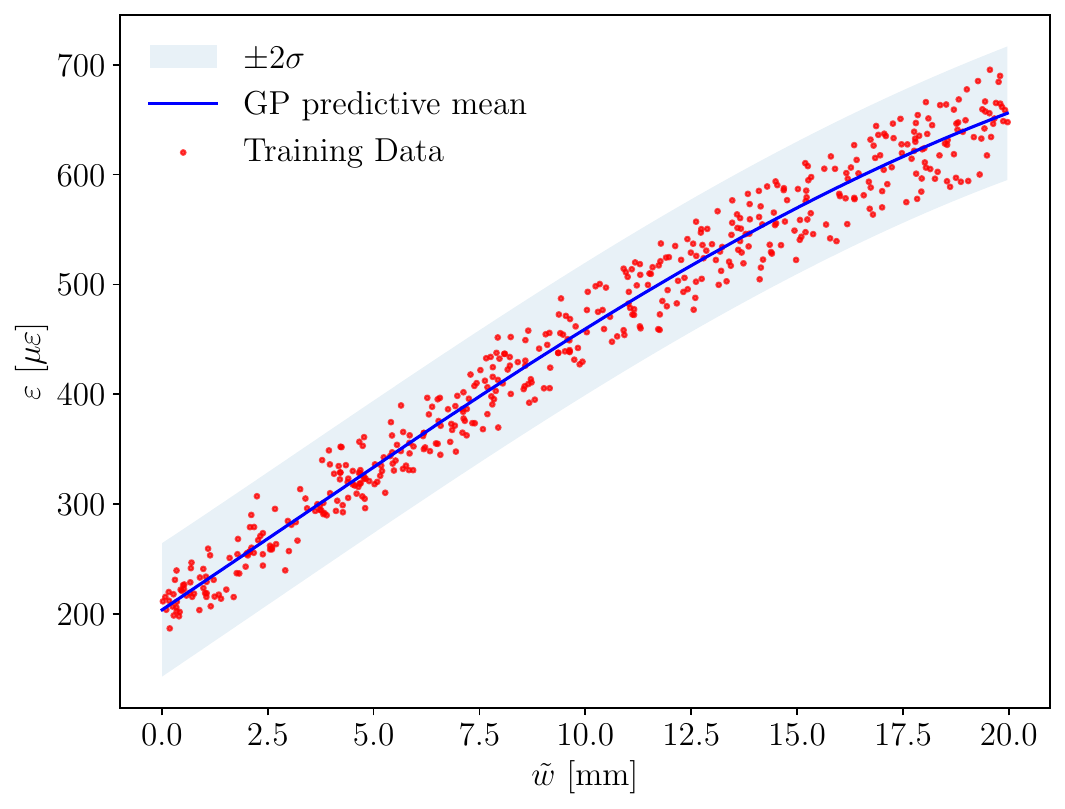}
    \caption{GPR surrogate model for Plate 6. Red markers represent the training-set (FE test data), the blue line shows the posterior predictive mean, and the shaded region indicates a $2 \sigma$ credible area. Note that for the construction of the likelihood function we use only the mean of the GPR.}
    \label{fig:gpr}
  \end{center}
\end{figure}

The surrogate model was chosen to be a standard Gaussian Process Regression~(GPR) model~\cite{rasmussen_2005} because it guarantees against overfitting and automatically returns confidence intervals associated with its predictive capacity. %
The former property is quite relevant in our case for robustly capturing the general pattern of the dataset, as the presence of multiple out-of-plane deflections contributes to some level of variability in the data.
The predictive variance provided by the GPR was also analyzed to assess the relative spread of strain values for different deflection amplitudes (i.e., the coefficient of variation), which was found to be $\sim 7 \%$ at the mean deflection amplitude. %
Given this level of variance, while acknowledging the computational complexity and inherent high-dimensionality that comes with hierarchical models, it was decided to only take the GPR mean into consideration in the likelihood function evaluation. %
\Cref{fig:gpr} illustrates a trained GPR model for a specific plate 
$k \in K$. %
Independent surrogate models were trained for each task to prevent bias in the mean prediction, and thus the model operator $\mathcal{M}(\cdot)$ in \Cref{eq:3} is more pertinently denoted as $\mathcal{M}_k(\cdot)$ for clarity.

\subsection{The hierarchical model}

This work compares a partially-pooled hierarchical model, which assumes correlation between plate-specific parameters, with an independent (no-pooling) model that does not, to motivate sharing information within the group of plates. %
Pooling refers to the way information is shared across members in hierarchical models; a no-pooling approach predicts separate parameters for each member, while conversely, a complete pooling approach assumes all members share the same parameters. %
Partial pooling strikes a balance by allowing some parameters to vary across members and correlation is enabled among them by conditioning on shared latent variables \cite{gelman_2013}. %
Since the deflection amplitudes related to each plate are assumed to be sampled from distributions with correlated parameters for the partially-pooled hierarchical model, one expects that the data-rich members ($k=1,\ldots,5$) should support the data-scarce one ($k=6$).

The hierarchical Bayesian model structure employed in this work is now presented. %
This follows from the work of Brealy et al. \cite{brealy_2024}. %
The likelihood function for the model was constructed using the commonly employed zero-mean Gaussian measurement error model,
\begin{equation}
    \left\{\left\{\varepsilon_{i,k} \right\}_{i=1}^{N_k}\right\}_{k=1}^K \sim
    \mathcal{N}(\mathcal{M}_k(\tilde{w}_{i,k}), \gamma^2)
    \label{eq:4}
\end{equation}

\noindent where $\gamma^2$ is the variance of the measurement error term $\epsilon$. %
This was considered to be the same for each plate under the assumption that the same hardware were used for data acquisition.

To start the Bayesian formulation, one can set prior distributions over the
deflection amplitude realizations $\tilde{w}_{i,k}$ related to each plate:
\begin{equation}
    \left\{\left\{\tilde{w}_{i,k} \right\}_{i=1}^{N_k}\right\}_{k=1}^K \sim
    \mathcal{N}(\mu_{\tilde{w}_{k}}, \sigma_{\tilde{w}_{k}}^2)
    \label{eq:5}
\end{equation}

The expectation $\mu_{\tilde{w}_{k}}$ and variance $\sigma_{\tilde{w}_{k}}^2$ associated with each plate are themselves sampled from global Normal distributions,
\begin{equation}
    \left\{\mu_{\tilde{w}_{k}}\right\}_{k=1}^K \sim
    \mathcal{N}(\mu_{\mu}, \sigma_{\mu}^2)
\end{equation}
\begin{equation}
    \left\{\sigma_{\tilde{w}_{k}}\right\}_{k=1}^K \sim
    \mathcal{N}(\mu_{\sigma}, \sigma_{\sigma}^2)
\end{equation}

\noindent where $\mu_{\mu}$, $\sigma_{\mu}^2$, $\mu_{\sigma}$ and $\sigma_{\sigma}^2$ are the higher-level parameters, which are shared among plates. %
One should also encode prior knowledge of the learning tasks as prior distributions over these higher-level parameters, referred to as hyperpriors. %
All four were assigned Gamma hyperpriors, more specifically:
\begin{equation}
    \mu_{\mu} \sim
    \mathcal{G}(\text{shape}=3, \text{rate}=0.2)
    \label{eq:8}
\end{equation}
\begin{equation}
    \sigma_{\mu} \sim
    \mathcal{G}(\text{shape}=0.8, \text{rate}=0.35)
    \label{eq:9}
\end{equation}
\begin{equation}
    \mu_{\sigma} \sim
    \mathcal{G}(\text{shape}=3.6, \text{rate}=6)
    \label{eq:10}
\end{equation}
\begin{equation}
    \sigma_{\sigma} \sim
    \mathcal{G}(\text{shape}=4.8, \text{rate}=16)
    \label{eq:11}
\end{equation}

The Gamma distribution was chosen to ensure positivity, which is essential for the physical parameters involved in the model. %
The constants in \Cref{eq:8,eq:9,eq:10,eq:11} (given in mm) were specified using the formulas for the two basic statistical moments (i.e., expected value and variance) of a Gamma-distributed random variable $\theta$. %
These are given by~\cite{gelman_2013},
\begin{equation}
    \text{E}[\theta] = \frac{\text{shape}} {\text{rate}}
\end{equation}
\begin{equation}
    \text{Var}[\theta] = \frac{\text{shape}} {\text{rate}^2}
\end{equation}

Following this, the shape and rate parameters were chosen in a way that the resulting variance for each higher-level parameter allows for sufficient exploration across a reasonable range of values. %
With the same rationale, the expected value was deliberately set to deviate to a certain extent from the ground-truth values used in data generation, ensuring that the model actually learns from the data. %
We also ensured that the chosen hyperprior parameters led to parameter/hyperparameter samples that, ultimately, determined non-negative deflection amplitudes, despite using Normal distributions.

Similarly to the higher-level parameters, a Gamma prior distribution was set for the shared noise variance $\gamma^2$,
\begin{equation}
    \gamma \sim
    \mathcal{G}(\text{shape}=80, \text{rate}=16)
\end{equation}

\noindent where the chosen parameter values reflect a fairly narrow support of possible values, as practically in most cases measurement noise is known a priori. %
Note that hypepriors for $\gamma$ are given in $\mu \varepsilon$.

\begin{figure}[ht]
  \begin{center}
    \includegraphics[width=0.95\textwidth]{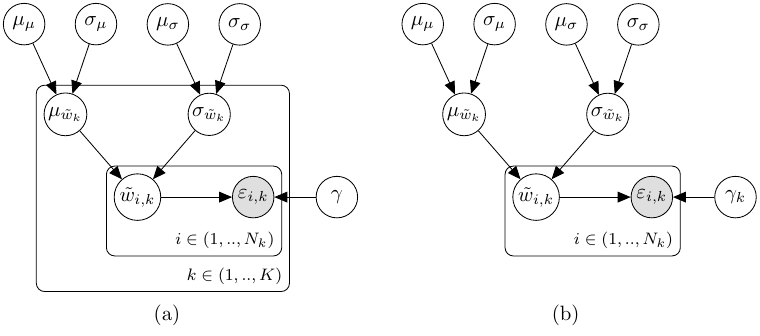}
    \caption{DAGs representing (a) the hierarchical (partial-pooling) model and (b) the independent model.}
    \label{fig:dags}
  \end{center}
\end{figure}

\Cref{fig:dags}(a) shows a directed acyclic graph (DAG) of the Bayesian model under a hierarchical approach. %
The plate-level parameters are indexed by $k$, and plate notation is used to indicate that these nodes are repeated over the $K$ plate components. %
Shaded and unshaded nodes represent observed and latent variables, respectively.  %
Higher-level nodes are positioned outside the plates and are not indexed by $k$, denoting that they are shared across all members. %
\Cref{fig:dags}(b) illustrates the DAG for the independent approach. %
The independent models can be derived by removing the 
$K$ plate from the hierarchical model's DAG and explicitly indexing the noise variance by $k$. %
Although the plate-level parameters are still conditioned on shared higher-level nodes, in this case, these nodes are only informed by the data from one specific member.

\section{RESULTS}\label{s:res}

\subsection{Inference}

Inference of the parameters was performed using MCMC methods, via the no U-turn variant (NUTS) of Hamiltonian Monte Carlo (HMC) \cite{hoffman_2014}. %
The implementation was carried out in {\fontfamily{qcr}\selectfont NumPyro} \cite{phan_2019}, 
which uses automatic differentiation to make gradient-based samplers like NUTS feasible.

\begin{figure}[!ht]
  \begin{center}
    \includegraphics[width=\textwidth]{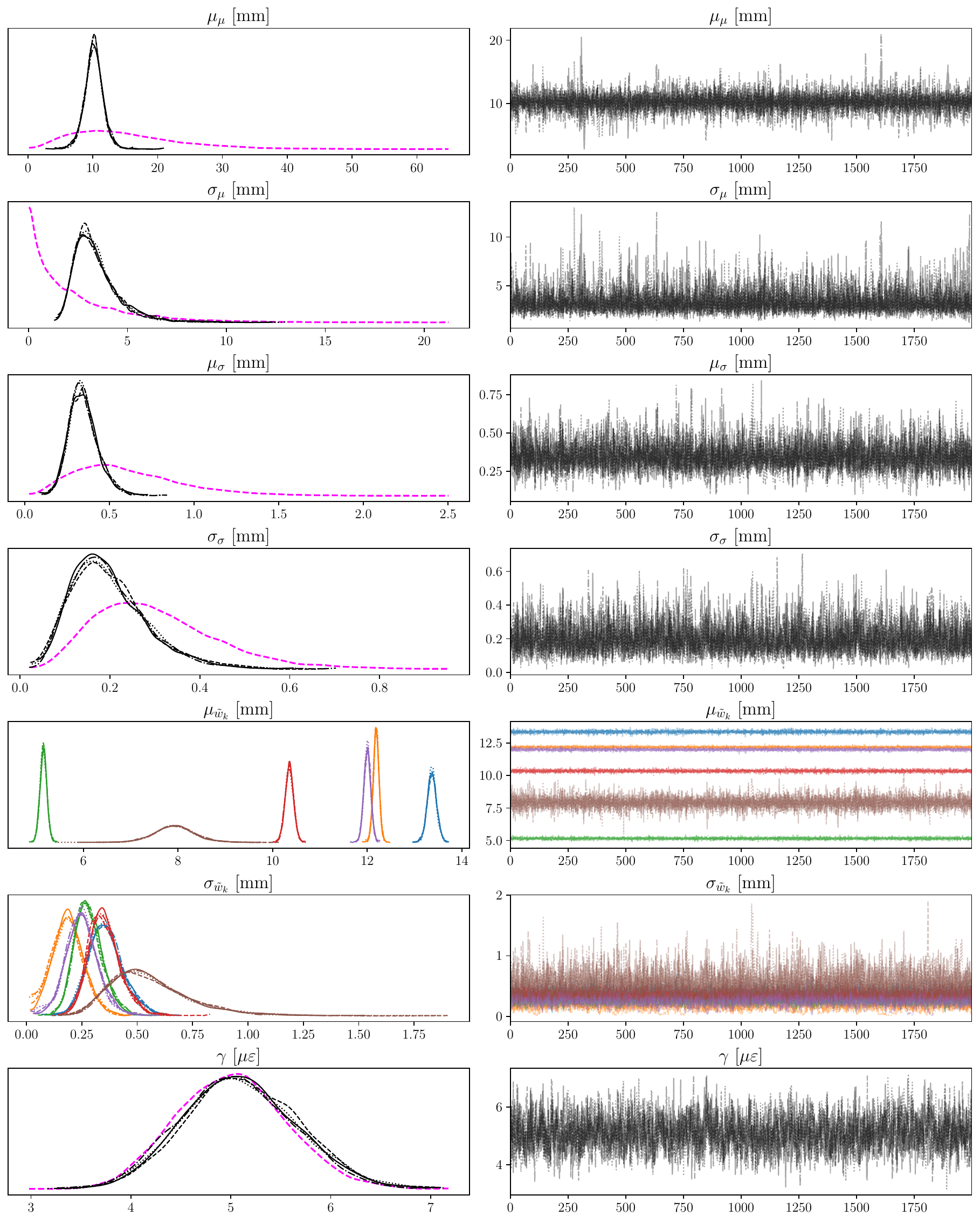}
    \caption{KDE-based posteriors of hierarchical model parameters/hyperparameters (left panel) and trace plots (right panel). Dashed magenta lines show the prior distributions applied to the shared parameters. The color scheme for $k=\{1,\ldots,6\}$ follows: blue, orange, green, red, purple, and brown.}
    \label{fig:traces}
  \end{center}
\end{figure}

The setup comprised of 4000 samples for warmup and 2000 samples for inference, run across four parallel chains, with each chain starting from a randomly generated initial state to ensure convergence. %
The rank-normalized $\hat{R}$ diagnostic \cite{vehtari_2021} was employed as quantitative check of convergence, and posteriors were accepted when this was lower than 1.01. %
This was satisfied for all parameters/hyperparameters, as can be seen by the trace plots on the right-hand side of \Cref{fig:traces}. %
The resulting kernel density function (KDE)-based posteriors for all chains (shown by each line style) are also shown on the left-hand side of the figure. %
The hyperpriors for the shared parameters have been plotted with magenta dashed lines, while the color scheme for the plate-specific parameters follows the convention used in \Cref{fig:clusters}.

From the results, one notices that the modes of the posterior distributions for the shared higher-level parameters (apart from $\gamma$) have shifted away from their prior modes. %
Additionally, the variance of these posterior distributions is notably reduced. %
These observations provide assurance that the algorithm is working properly and that the model has learned from the measurements. %
As expected, the posterior distribution of the noise variance 
$\gamma$ remains relatively unchanged from its prior due to the tightly constrained prior range and the inherent properties of the Gamma distribution, which features shallow tails that might affect the NUTS sampler during exploration of the posterior. %
However, this is compensated via $\mu_{\sigma}$ which encodes any additional variance observed in the data, and behaves as intended. 

For the plate-specific parameters 
$\mu_{\tilde{w}_{k}}$ and $\sigma_{\tilde{w}_{k}}$, the posterior variance for the data-scarce plate (i.e., $k=6$) is considerably larger than that of the data-rich plates. %
This is expected, given that it is informed by only two observations, whereas the other plates have 20 data points each. %
A comparison was conducted between this posterior distribution and the one obtained when modeling this plate independently (i.e., without pooling information from the broader dataset) in order to show the benefit of hierarchical models.

\subsection{Damage detection}

While the inferred latent variables related to deflection levels are those that describe the underlying structural behavior, they are not directly observed in an operational setting. %
Instead, for downstream tasks, such as damage detection, one should revert to the strain space, as this is what can be monitored and used by operators. 

Along these lines, the posterior distributions of the plate-specific parameters were used for damage detection in the following way. %
First, posterior MCMC samples were taken for the expectation and variance for plate $k=6$, to draw a deflection amplitude each time from a Normal distribution with the same statistics as the parameter draws (according to \Cref{eq:5}). %
Next, to obtain posterior predictive samples for the transverse strain, the drawn deflection amplitude samples were evaluated through the mean function of $\mathcal{M}_{k=6}(\cdot)$, and the likelihood function was applied using the posterior predictive measurement noise $\gamma^2$ (see \Cref{eq:4}), for each sample. %
KDE estimation was then employed to approximate the posterior predictive distribution of strains, using the samples from the likelihood. %
This process was applied for both the hierarchical (partial-pooling) and the independent (no-pooling) approaches.

\begin{figure}[ht]
  \begin{center}
    \includegraphics[width=10cm]{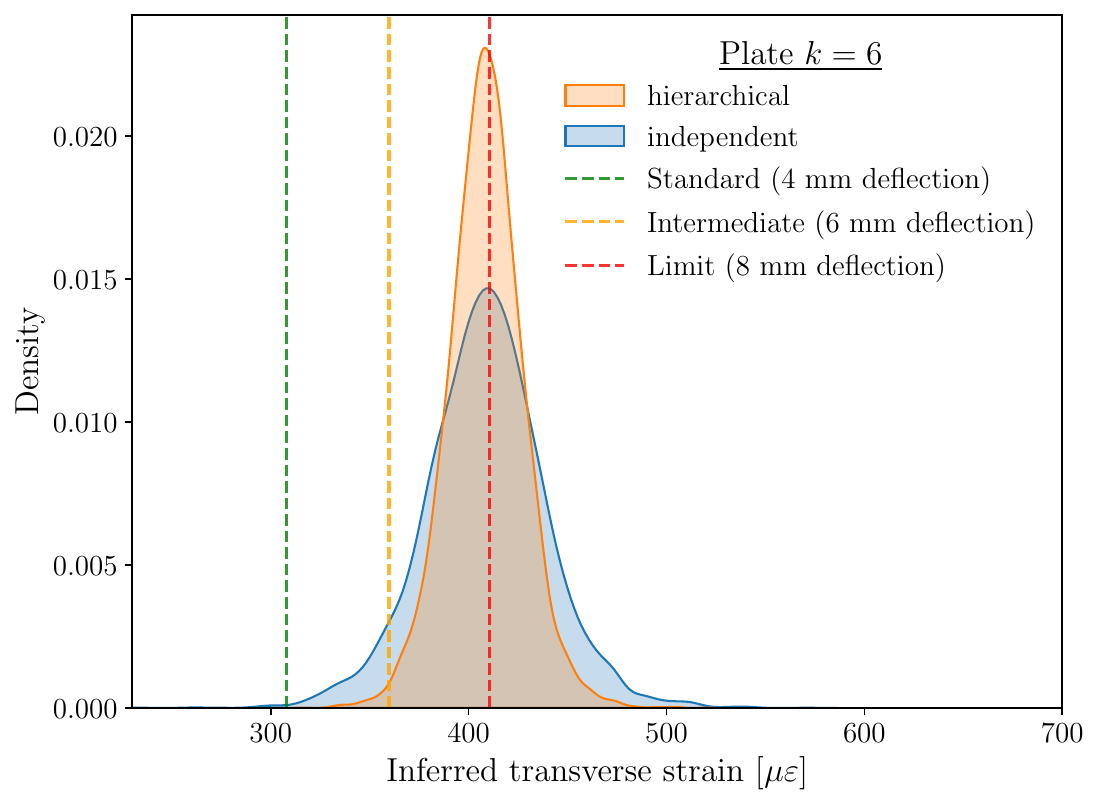}
    \caption{KDE-based posterior predictive densities of strain under a hierarchical (partial-pooling) approach and an independent (no-pooling) approach. The dashed lines represent averaged strain values for different deflection levels.}
    \label{fig:hbm_vs_independent}
  \end{center}
\end{figure}

The posterior predictive distributions for the transverse strain are shown in \Cref{fig:hbm_vs_independent}, with the vertical lines indicating strain values associated with different deflection thresholds set by maritime regulatory bodies. %
Namely, the 8 mm level corresponds to a limit deflection threshold that can be seen as a damage indicator, while the 4 mm represents a standard level expected under normal operating conditions. %
The 6 mm level constitutes an intermediate state. %
The strain values for these thresholds were computed by passing them into the surrogate model an equal number of times as the posterior MCMC samples of $\gamma^2$, drawing observations as per the likelihood, and then averaging the results for each deflection level.

It is evident that the partial pooling model has managed a noticeable reduction in the level of uncertainty in its predictions compared to the no-pooling model. %
This follows from the fact that the parameters pertaining to the latent variable have been learned within a joint inference setting. %
Such a reduction in uncertainty is particularly valuable in industrial applications, where high uncertainty can lead to operational challenges. %
For instance, if a deflection amplitude exceeding the 8 mm threshold implies inspection, large uncertainties could result in higher probability of false alarms to occur. %
This, in turn, would lead to unnecessary downtime and increased costs for stakeholders. %
At the same time, within a signal detection theory context \cite{kay_1998}, the reduction in uncertainty enhances the probability of detection by improving the model’s ability to distinguish actual damage from normal variations. %
This leads to a more reliable assessment of structural integrity and reduces the risk of missed detections.

\section{CONCLUSIONS}\label{s:concls}

In this work, a hierarchical Bayesian model has been constructed to learn the latent (both local and global) parameters that drive a fabrication-induced deterioration phenomenon commonly encountered in ship hull structures, i.e., out-of-plane deflections. %
A set of deflected plates was assumed in a region of the double-bottom structure of a containership, and synthetic strain observations were generated through an FE model. %
These observations served as features in the inference process, where an MCMC sampler, specifically the NUTS algorithm, was employed. %
Given the computational demands of NUTS, a surrogate model was used in place of the FE model to improve computational efficiency.

To demonstrate the capabilities of hierarchical modeling, a specific plate was assumed with less data compared to the others. %
The hierarchical model enabled this data-poor plate to borrow statistical strength from the broader population, reducing prediction uncertainty. %
The refined posterior uncertainties of the parameters were compared against an independent no-pooling approach, with results showing that the hierarchical model enables more informed decision-making in data sparse scenarios.

\bibliographystyle{unsrtnatemph}

\bibliography{ref}

\end{document}